\documentclass[useAMS,usenatbib]{mn2e}
\usepackage{epsfig}

\def\com{{\rm com}}

\def\p{{\rm peak}}
\def\iso{{\rm iso}}

\def\bol{{\rm bol}}

\def\i{{\rm i}}

\def\sw{{\em Swift}}
\def\gl{{\em GLAST}}

\title[Variation of the Amati Relation with Redshift] 
{Variation of the Amati Relation with the Cosmological Redshift: a Selection
Effect or an Evolution Effect?}
\author[Li-Xin Li]{Li-Xin Li\thanks{E-mail: lxl@mpa-garching.mpg.de}\\
Max-Planck-Institut f\"ur Astrophysik, 85741 Garching, Germany}

\begin{document}


\date{Accepted 2007 April 23.  Received 2007 April 21; in original form 2007 April 12}

\pagerange{\pageref{firstpage}--\pageref{lastpage}} \pubyear{2006}

\maketitle

\label{firstpage}

\begin{abstract}
Because of the limit in the number of gamma-ray bursts (GRBs) with available
redshifts and spectra, all current investigations on the correlation among
GRB variables use burst samples with redshifts that span a very large range. 
The evolution and selection effects have thus been ignored, which might
have important influence on the results. In this Letter, we divide the 48
long-duration GRBs in Amati (2006, 2007) into four groups with redshift from 
low to high, each group contains 12 GRBs. Then we fit each group with the
Amati relation $\log E_\iso = a + b \log E_\p$, and check if the parameters 
$a$ and $b$ evolve with the GRB redshift. We find that $a$ and $b$ vary with
the mean redshift of the GRBs in each group systematically and significantly. 
Monte-Carlo simulations show that there is only $\sim 4$ percent of chance 
that the variation is caused by the selection effect arising from the fluence
limit. Hence, our results may indicate that GRBs evolve strongly with the 
cosmological redshift.
\end{abstract}


\section{Introduction}
\label{intro}

A remarkable progress in the observation of gamma-ray bursts (GRBs) has
been the identification of several very good correlations among the GRB
observables (see Schaefer 2007 for a review). Based on several of those
correlations, some people have eagerly proposed to use GRBs as standard
candles to probe the cosmological Hubble diagram to very high redshift
\citep[and references therein]{sch03,dai04,ghi04,lam05,fir06,sch07}.
Enlightening comments and criticism on GRBs as standard candles can be
found in \citet{blo03} and \citet{fri05}.

All the GRB correlations have been obtained by fitting a hybrid GRB sample 
without discriminating the redshift. Indeed, the redshift in the sample 
usually spans a very large range: from $z\sim 0.1$ up to $z \sim 6$. This is 
of course caused by the fact that we do not have an enough number of GRBs 
with measured redshifts limited in a small range. Then, inevitably, the 
effect of the GRB evolution with the redshift, and the selection effects, 
have been ignored. This raises an important question about whether the 
relations that people have found reflect the true physics of GRBs or they 
are just superficial. 
[See \citet{ban05} for a nice discussion on the selection effect and the 
correlation between the GRB peak spectral energy and the 
isotropic-equivalent/jet collimated energy.]

For objects distributing from $z\sim 0.1$ to $z\sim 6$, it is hard to 
believe that they do not evolve. There are cumulative evidences suggesting
that long-duration GRBs prefer to occur in low-metallicity galaxies
\citep{fyn03,hjo03,lef03,sol05,fru06,sta06}. With a sample of five nearby 
GRBs, \citet{sta06} have found that the isotropic energy of GRBs is
anti-correlated with the metallicity in the host galaxy \citep[see, however,] 
[]{wol07}. It is well known that metallicities evolve strongly with
the cosmological redshift \citep{kew05,sav05}. Hence, the evolution of GRBs 
with the redshift is naturally expected \citep[see, e.g.,][]{lan06}.

In this Letter, we use the Amati relation as an example to test the cosmic
evolution of GRBs. The Amati relation is a correlation between the
isotropic-equivalent energy of long-duration GRBs and the peak energy of 
their integrated spectra in the GRB frame \citep{ama02}
\begin{eqnarray}
        \log E_\iso = a + b \log E_\p \;. \label{amati}
\end{eqnarray}
The isotropic-equivalent energy $E_\iso$ is defined in the $1$--$10000$ 
keV band in the GRB frame.

With a sample of 41 long GRBs with firmly determined redshifts and peak 
spectral energy, \citet{ama06} has obtained that $a= -3.35$ and $b= 1.75$ 
with the least squares method ($E_\p$ in keV and $E_\iso$ in $10^{52}$ erg); 
and $a= -4.04$ and $b= 2.04$ with the maximum likelihood method with an 
intrinsic dispersion in the relation (\ref{amati}) being included. Long GRBs 
detected by \sw\, and having measured redshifts and $E_\p$ are found to be 
consistent with the Amati relation \citep{ama07}.

The difference in the values of the parameters obtained with the two methods 
can be explained as follows. The maximum likelihood method directly probes 
the intrinsic relation between the two variables, $x=\log E_\iso$ and $y = 
\log E_\p$ \citep{dag05}. However, roughly speaking, the least squares method 
estimates the average value of $x$ at a given $y$, $\langle x\rangle =
a^\prime + b^\prime y$. \citet{tee84} has shown that, when $x$ has a Gaussian 
distribution with a dispersion $\sigma_x$, and the relation $x =a + b y$ has 
an intrinsic dispersion $\sigma_y^\i$ in $y$, $b^\prime$ is related to $b$ by 
\begin{eqnarray}
	b^\prime = b \left(1+\frac{b^2{\sigma_y^\i}^2}{\sigma_x^2}
		\right)^{-1} \;.
	\label{tee}
\end{eqnarray}
\citet{ama06} has found that $\sigma_x\approx 0.9$, $b\approx 2.04$, and 
$\sigma_y^\i\approx 0.15$. Then by equation (\ref{tee}) we have $b^\prime
\approx 1.83$, which is close to the value of $1.75$ obtained by the least 
squares method.

To test if the Amati relation varies with the cosmological redshift, in this 
Letter we separate a sample of 48 long GRBs [consisting of the long 41 GRBs  
from \citet{ama06} and seven additional \sw\, long GRBs from \citet{ama07}] 
into four groups by the GRB redshift. That is, we sort the GRBs by their 
redshifts, and divide them into four groups with redshifts distributing from 
low values to high values. Each group contains 12 GRBs (for details see 
Section \ref{evol}). We then fit each group by equation (\ref{amati}) and 
calculate the mean redshift, and check if the values of $a$ and $b$ 
evolve with the redshift.

As we will see that, the values of $a$ and $b$ strongly vary with the
redshift. The variation is not likely to arise from the selection effect and
hence may indicate that GRBs evolve strongly with the cosmological redshift.

Throughout the Letter, we follow \citet{ama06} to adopt a cosmology with 
$\Omega_m = 0.3$, $\Omega_\Lambda=0.7$, and $H_0 = 70$ km s$^{-1}$ Mpc$^{-1}$.

\section{Variation of the Amati Relation with the Cosmological Redshift}
\label{evol}

To test if the Amati relation (\ref{amati}) evolves with the redshift, we
separate a sample of 48 long GRBs into four groups according to the redshift 
of the GRBs, then fit each group with equation (\ref{amati}).

The sample contains 41 long GRBs from \citet[Table 1]{ama06}, and seven 
additional \sw\, long GRBs from \citet{ama07}. The additional seven \sw\, 
GRBs are 060115, 060124, 060206, 060418, 060707, 060927, and 061007. Since 
a GRB sample with $z \la 0.1$ is very incomplete, we select only GRBs with 
$z>0.1$ and hence GRB 060218 ($z=0.0331$) is not included. GRB 060614 is 
also excluded from our sample because of the very large uncertainty in its
$E_\p$ \citep{ama07b}.

A least squares fit to the 48 GRBs as a single sample with equation 
(\ref{amati}) leads to $a= -3.42$, $b= 1.78$, with $\chi_r^2 = 5.9$. The 
$\chi^2_r$ is the reduced $\chi^2$, i.e., the $\chi^2$ of the fit divided 
by the degree of freedom. A maximum likelihood fit, which includes an
intrinsic dispersion $\sigma_\i$ in $\log E_\p$ in the relation (\ref{amati}),
leads to $a=-4.08$, $b=2.04$, and $\sigma_\i = 0.14$. These results are 
consistent with that obtained with 41 GRBs by \citet{ama06}.

The redshift of the 48 GRBs spans a range of $0.17$--$5.6$. GRB 030329 has 
the minimum redshift ($z=0.17$). GRB 060927 has the maximum redshift ($z=
5.6$). The mean redshift is $\langle z\rangle = 1.685$. We separate the 
48 GRBs into four groups with redshifts from low to high, each group contains
12 GRBs:\\
{\bf Group A}---12 GRBs, $0.1< z <0.84$, $\langle z\rangle = 0.56$;\\
{\bf Group B}---12 GRBs, $0.84\le z <1.3$, $\langle z\rangle = 1.02$;\\
{\bf Group C}---12 GRBs, $1.3\le z <2.3$, $\langle z\rangle = 1.76$;\\
{\bf Group D}---12 GRBs, $2.3\le z \le 5.6$, $\langle z\rangle = 3.40$.

\begin{figure}
\vspace{2pt}
\includegraphics[angle=0,scale=0.48]{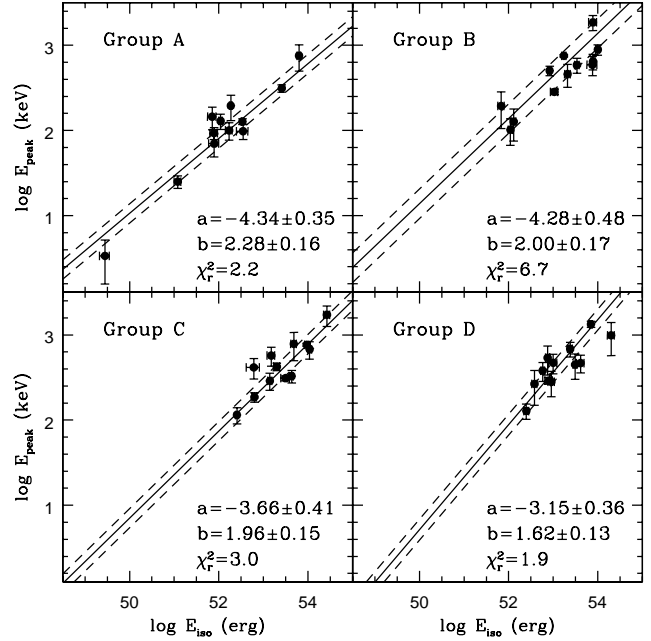}
\caption{Least squares fit to each of the four groups of GRBs (data points 
with error bars; see text) by equation (\ref{amati}) (solid line). The two 
dashed lines mark the 1-$\sigma$ deviation of the fit.
}
\label{ee_z2}
\end{figure}

The least squares fit to each group of GRBs by equation (\ref{amati}), taking
into account the errors in both $E_\p$ and $E_\iso$, is shown in 
Fig.~\ref{ee_z2}. Immediately one can see that, except Group B, the $\chi_r^2$
for each group is smaller than that obtained by fitting the whole sample of 
GRBs. This fact indicates that treating the GRBs at different redshifts as a 
single sample may increase the data dispersion (see Fig.~\ref{ee_s} below).

\begin{figure}
\vspace{2pt}
\includegraphics[angle=0,scale=0.467]{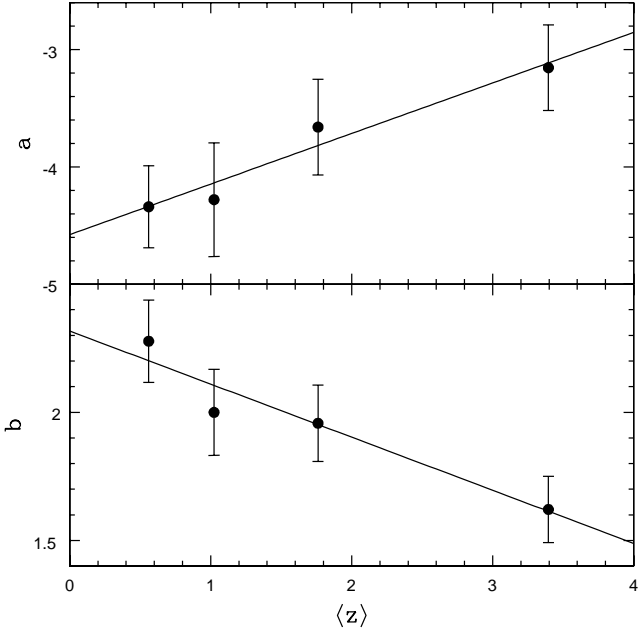}
\caption{The fitted values of $a$ and $b$ against the mean redshift of 
GRBs. Each data point with error bars represents a group of GRBs (A, B, 
C, and D). The solid line is a least squares linear fit to $a$--$\langle 
z\rangle$ and $b$--$\langle z\rangle$.
}
\label{ee_ab}
\end{figure}

We find that the values of $a$ and $b$ vary with the mean redshift of the 
GRBs monotonically.\footnote{Generally, the variations of $a$ and $b$ are 
not independent, see e.g., Li \& Paczy\'nski 2006} In Fig.~\ref{ee_ab} we 
plot $a$ and $b$ against $\langle z\rangle$. Clearly, $a$ and $b$ are 
correlated/anti-correlated with $\langle z\rangle$. The Pearson linear 
correlation coefficient between $a$ and $\langle z\rangle$ is $r(a,\langle 
z\rangle) = 0.975$, corresponding to a probability $P = 0.025$ for a zero 
correlation. The correlation coefficient between $b$ and $\langle z\rangle$ 
is $r(b,\langle z\rangle) = -0.960$, corresponding to a probability $P = 
0.040$ for a zero correlation.

A least squares linear fit to $a$--$\langle z\rangle$ (the solid line in the 
upper panel of Fig.~\ref{ee_ab}) leads to
\begin{eqnarray}
	a = -4.58 (\pm 0.36) + 0.43 (\pm 0.17)\, z \;, \label{az_f}
\end{eqnarray}
with $\chi^2_r = 0.13$. A least squares linear fit to $b$--$\langle z\rangle$
(the solid line in the lower panel of Fig.~\ref{ee_ab}) leads to
\begin{eqnarray}
	b = 2.32 (\pm 0.15) - 0.207 (\pm 0.066)\, z \;, \label{bz_f}
\end{eqnarray}
with $\chi^2_r = 0.31$.

The results indicate that $a$ and $b$ strongly evolve with the cosmological
redshift.

\begin{figure}
\vspace{2pt}
\includegraphics[angle=0,scale=0.465]{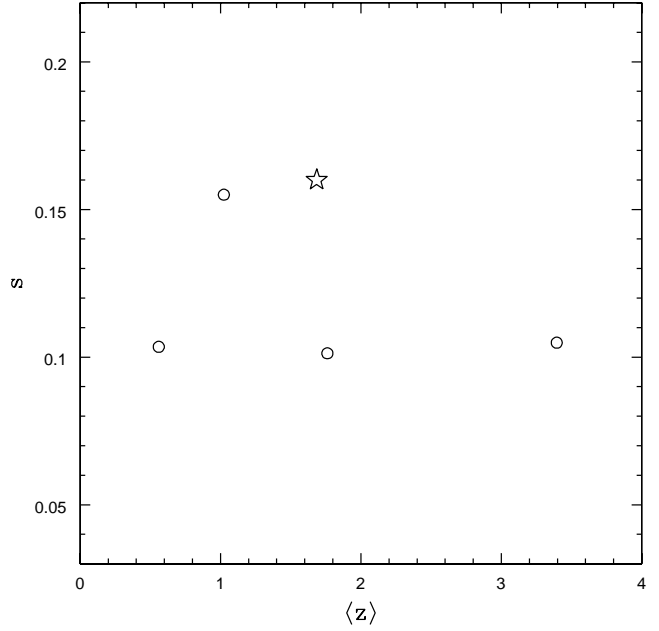}
\caption{The deviation of fit. Each circle corresponds to a group of GRBs.
The star represents the result obtained by fitting the whole sample (48 GRBs),
which is $s=0.16$.
}
\label{ee_s}
\end{figure}

In Fig.~\ref{ee_s} we plot the deviation of fit ($s$; see Bevington \& 
Robinson 1992, Li \& Paczy\'nski 2006) against the mean redshift of GRBs. 
There is not a clear trend for $s$ to vary with $\langle z\rangle$. But it 
appears that the deviation of fit of each group is smaller than that of the 
whole sample.

\section{Is the Variation Caused by the Selection Effect?}
\label{selection}

To check if the variation of $a$ and $b$ with the cosmological redshift is
caused by the selection effect, we use Monte-Carlo simulations to generate
a sample of GRBs according to a pre-assumed Amati relation (\ref{amati})
and with a limit in the observed GRB fluence. Then, we divide the sample into
four groups by the GRB redshift and fit each group by equation (\ref{amati}),
just as we did in Section \ref{evol}.

The lower limit in the bolometric fluence, $F_{\bol,\lim}$, leads to a lower
limit in the isotropic-equivalent energy of a detectable burst at redshift 
$z$ 
\begin{eqnarray}
	E_{\iso,\lim} = 4\pi D_\com^2(1+z)F_{\bol,\lim} \;,
	\label{limit}
\end{eqnarray}
where $D_\com$ is the comoving distance to the burst.

\begin{figure}
\vspace{2pt}
\includegraphics[angle=0,scale=0.47]{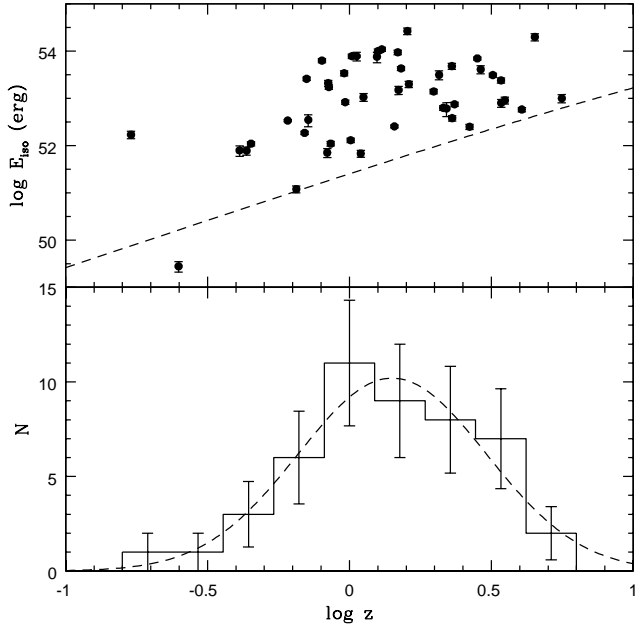}
\caption{Upper panel: The isotropic-equivalent energy versus the redshift 
for the 48 GRBs in the sample. The dashed line is the limit given by 
equation (\ref{limit}) with $F_{\bol,\lim} = 1.2\times 10^{-6}$ erg cm$^{-2}$.
Lower panel: The redshift distribution for the GRBs (histogram). The vertical 
error bar represents the Poisson fluctuation. The dashed curve is a fit to 
the $N$--$\log z$ relation by a Gaussian function. The bin size in $\log z$ is 
$0.1778$.
}
\label{eiso_z2}
\end{figure}

In Fig.~\ref{eiso_z2} upper panel, we plot the isotropic energy of the 48 GRBs 
in the sample of \citet{ama06,ama07} against their redshifts. The isotropic 
energy is clearly correlated with the redshift, with a Pearson linear 
correlation coefficient $r = 0.437$ and a probability $P = 0.0019$ for a zero 
correlation. The dashed line in the figure is the limit given by equation 
(\ref{limit}) with $F_{\bol,\lim} = 1.2\times 10^{-6}$ erg cm$^{-2}$, which 
reasonably represents the selection effect.

The distribution of the redshifts of the GRBs in the sample is plotted in
Fig.~\ref{eiso_z2} lower panel. It can be fitted by a log-normal distribution,
with a mean $\mu= 0.151$ and a dispersion $\sigma= 0.332$ in $\log z$. The 
$\chi_r^2$ of the fit is $0.25$. Then, the frequency distribution in $\log z$ 
is
\begin{eqnarray}
	f_1(\log z) = \frac{1}{\sqrt{2\pi}\sigma} \exp\left[-\frac{(\log z
		-\mu)^2}{2\sigma^2}\right] \;,
	\label{f_logz}
\end{eqnarray}
whose integration over $\log z$ (from $-\infty$ to $\infty$) is unity.

The distribution of the isotropic-equivalent energy is also described by a
log-normal distribution, with a mean $= 1.09$ and a dispersion $=0.85$ in 
$\log E_\iso$ ($E_\iso$ in $10^{52}$ erg).

Define $x\equiv \log E_\iso$ and $y\equiv \log E_\p$, where $E_\iso$ is in
$10^{52}$ erg, and $E_\p$ is in keV. Assuming that the Amati relation is 
valid and independent of the redshift, and for a given $x$ we have $y = m x 
+p$ with an intrinsic dispersion $\sigma_\i$ in $y$. Then, for a given $x$, 
the Gaussian distribution of $y$ is given by
\begin{eqnarray}
	f_2(y) = \frac{1}{\sqrt{2\pi}\sigma_\i} \exp\left[-\frac{(y-m x
		-p)^2}{2\sigma_\i^2}\right] \;.
	\label{f_y}
\end{eqnarray}
By our maximum likelihood fit results in Section \ref{evol}, we take $m=0.49$,
$p = 2.00$, and $\sigma_\i = 0.14$.

The Monte-Carlo simulation is done as follows. First, we generate $N$
redshifts with the distribution in equation (\ref{f_logz}). Then, at each
redshift, we generate an isotropic-equivalent energy with a log-normal
distribution (mean of $\log E_\iso = 1.09$, dispersion $= 0.85$) and 
satisfying $E_\iso > E_{\iso,\lim}$. Finally, for any pair of $(z, E_\iso)$,
we generate a peak energy $E_\p$ according to the distribution in equation
(\ref{f_y}). Then we have a sample of $N$ GRBs, each GRB has a redshift, a 
peak spectral energy, and an isotropic-equivalent energy. These GRBs satisfy 
the distributions described above, and the select condition defined by 
equation (\ref{limit}) (with $F_{\bol,\lim} = 1.2\times 10^{-6}$ erg 
cm$^{-2}$).

With the above approach, we generated $N=4000$ GRBs. We divided them into
four groups by redshift, and each group contains 1000 GRBs. Then, we fitted 
each group of GRBs by equation (\ref{amati}) and got the values of $a$ and 
$b$, and checked the evolution of $a$ and $b$ with the mean redshift $\langle 
z\rangle$. We repeated the process 10000 times, each time with a different 
sample of 4000 GRBs. We found that $a$ and $b$ indeed varied with $\langle 
z\rangle$
\begin{eqnarray}
	a = -3.57+0.105\, z \;, \hspace{1cm}
	b = 1.84 - 0.0347\, z \;.  \label{ab_z}
\end{eqnarray}

This variation was caused by the selection effect, i.e. the limit in
equation (\ref{limit}). If we turned off the limit, we found that $a$ and $b$ 
did not evolve with $\langle z\rangle$. However, comparing equation 
(\ref{ab_z}) to equations (\ref{az_f}) and (\ref{bz_f}), we found that the 
selection effect is not likely the cause for the evolution in equations 
(\ref{az_f}) and (\ref{bz_f}), since the $a$ and $b$ in equation (\ref{ab_z}) 
evolve too slowly with $z$. Even if we increased the value of $F_{\bol,\lim}$ 
to $10^{-5}$ erg cm$^{-2}$, we got $da/dz = 0.22$ and $db/dz = -0.065$, whose 
values are still too small to explain the evolution in equations (\ref{az_f}) 
and (\ref{bz_f}).

\begin{figure}
\vspace{2pt}
\includegraphics[angle=0,scale=0.457]{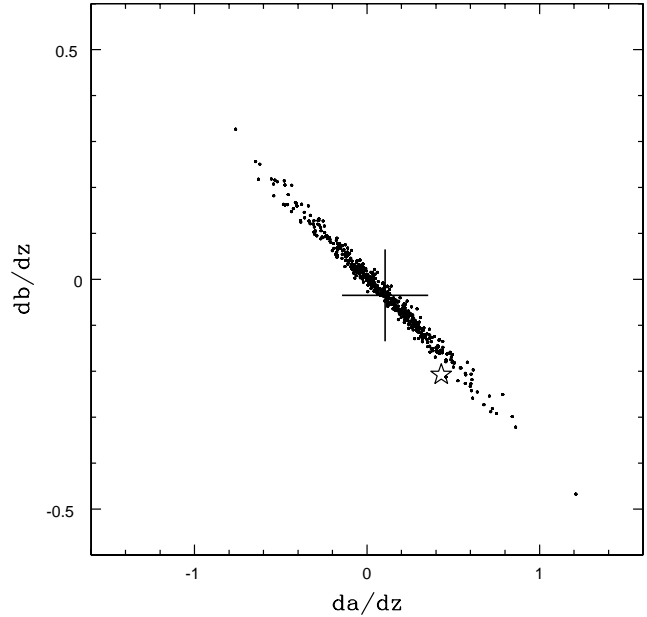}
\caption{The slopes $da/dz$ and $db/dz$ obtained from Monte-Carlo
simulations (points; 500 runs of 48 GRBs). The cross marks the average 
values of $da/dz = 0.105$ and $db/dz = -0.0347$, obtained with 10000 runs of 
4000 GRBs (see text). The star is the value obtained with the 48 observed 
GRBs (eqs.~\ref{az_f} and \ref{bz_f}). 
}
\label{ab_fit2}
\end{figure}

Of course, equation (\ref{ab_z}) only describes the average evolution of
$a$ and $b$ for the 10000 runs. For each run, the evolution may deviate
from equation (\ref{ab_z}). To obtain the chance probability for the
evolution in equations (\ref{az_f}) and (\ref{bz_f}) to arise from the 
selection effect, we used the Monte-Carlo simulation described above to
generate $N=48$ GRBs and repeated the process 500 times. For each 48 GRBs 
obtained in each run, we separate them into four groups and calculate 
$da/dz$ and $db/dz$ just as we did to the 10000 runs of 4000 GRBs. The 
results are shown in Fig.~\ref{ab_fit2}.

Based on our simulations (500 runs of 48 GRBs), we found that the probability 
for getting a pair of ($da/dz$,$db/dz$) with $da/dz >0.43$ and $db/dz <-0.207$
is $0.04$. Hence, we have only $\sim 4$ percent of chance that the variation 
presented in Section \ref{evol} is caused by the selection effect.

\section{Conclusions}
\label{concl}

If GRBs do not evolve with the redshift and the selection effects are not
important, we would expect that the Amati relation does not change with the 
redshift. Hence, from the variation of the Amati relation with the redshift
we may get some clues on the cosmic evolution of GRBs.

By dividing the 48 GRBs in \citet{ama06,ama07} into four groups by their 
redshifts and fitting each group separately, we have found that the 
isotropic-equivalent energy and the peak spectral energy of GRBs remain
being correlated in each group, even with a smaller dispersion than that
for the whole sample. However, the parameters $a$ and $b$ in the Amati 
relation (\ref{amati}) evolve strongly with the redshift (eqs.~\ref{az_f} 
and \ref{bz_f}).

Although the selection effect arising from the limit in the GRB fluence may 
cause a similar variation of $a$ and $b$ (eq.~\ref{ab_z}), generally the 
variation is too slow to explain what we have found for the observed GRBs
(eqs.~\ref{az_f} and \ref{bz_f}). With Monte-Carlo simulations we have shown 
that there is only $\sim 4$ percent of chance that the observed variation 
is caused by the selection effect. Hence, the variation of the Amati relation
with the redshift that we have discovered may reflect the cosmic evolution
of GRBs and indicates that GRBs are not standard candles.

Our results are limited by the small number of GRBs in the sample: we have
48 GRBs in total, and only $12$ GRBs in each group. To get a more 
reliable conclusion, the number of GRBs with well determined redshifts and 
spectra need be significantly expanded. Since the launch of \sw, the
fraction of GRBs with measured redshifts has increased rapidly. However, 
unfortunately, due to the narrow energy range of the Burst Alert Telescope 
(BAT) on \sw, the fraction of bursts that have accurately determined 
peak/isotropic energy has not increased proportionally. The Gamma-ray
Large Area Space Telescope (\gl) scheduled for launch in late 2007 will
provide us with more promise for this purpose \citep{omo06}.

We must also stress that our treatment on the selection effect has been
greatly simplified. The GRBs in the sample were detected and measured by 
different instruments, hence the selection effect is much more complicated. 
A more careful consideration of the various selection biases is required 
to determine if the observed evolution of the Amati relation reflects the 
cosmic evolution of GRBs.

No matter what the conclusion will be (the variation of parameters is caused
by the GRB evolution effect or by the selection effect), our results suggest
that it is a great risk to use GRBs with redshifts spanning a large range
as a single sample to draw physics by statistically analyzing the correlations
among observables. Although we have only tested the Amati relation, it would 
be surprising if any of the other relations \citep[e.g., the Ghirlanda 
relation;][]{ghi04a} does not change with the redshift.

\section*{Acknowledgments}

The author thanks the referee Dr. P. O'Brien for a very helpful report.
The Letter was based on a presentation by the author at the debate on 
``Through GRBs to $\Omega$ and $\Lambda?$'' during the conference
``070228: The Next Decade of GRB Afterglows'' held in Amsterdam, 19--23 
March 2007. The author acknowledges all the attendants at the debate 
for exciting and inspiring discussions.

\bsp

\label{lastpage}

\end{document}